\documentclass[aps,prl,twocolumn,groupedaddress,showpacs]{revtex4-1}
\usepackage{enumerate}
\usepackage{amsmath, amsthm, amssymb}
\usepackage{color,calc,graphicx,subfig}

\begin{document}

%\preprint{}

%\begin{CJK*}{UTF8}{}

\title{A simple protocol for certifying graph states and applications in quantum networks}



%\thanks{}
\author{Damian Markham}
\affiliation{Laboratoire d'Informatique de Paris 6, CNRS, UPMC-Sorbonne Universites, 4 place Jussieu, 75005 Paris, France}

\author{Alexandra Krause}
%\email[]{damian.markham@upmc.fr}
\affiliation{Laboratoire d'Informatique de Paris 6, CNRS, UPMC-Sorbonne Universites, 4 place Jussieu, 75005 Paris, France}
\affiliation{Freie Universität Berlin, 14195 Berlin, Germany}

%\author{Zizhu Wang (\CJKfamily{gbsn}王子竹)}
%\email[]{zizhu.wang@telecom-paristech.fr}


%\date{\today}

\begin{abstract}

We present a simple protocol for certifying graph states in quantum networks using stabiliser measurements.  The certification statements can easily be applied to different protocols using graph states. We see for example how it can be used to for measurement based verified quantum computation, certified sampling of random unitaries and quantum metrology and sharing quantum secrets over untrusted channels.
\end{abstract}

\maketitle
%\end{CJK*}

 \section{Introduction}
 
 Graph states are a family of multipartite quantum states, defined in one to one correspondance with a simple graph~\cite{HEB04}. They are incredibly useful resources across quantum information, acting as the key entanglement resource for error correction~\cite{SW01}, measurement based quantum computation~\cite{RB01}, quantum secret sharing~\cite{MS08} and more~\cite{HEB04}. Furthermore, they can be implemented in many different ways, for example in optics \cite{WL+16,BK+12,BM+14}, \cite{CY+17,YU+13} including on chip \cite{CO+16}, in ion traps~\cite{BM+11,MS+11}, super conducting qubits~\cite{SX+17} and NV centres~\cite{CK+16}. 
 
Many methods exist for testing graph states varying in the trust that must be assumed and the kind of statements that are made. 
With respect to trust assumptions, on the one hand techniques such as tomography~\cite{dAPS03} and entanglement witnesses~\cite{JMO11} make assumptions about the source and measurements (essentially that they are honest but noisy).  
On the other hand tests which require the least trust, where neither the source nor the measurement devices are trusted, such as self testing~\cite{M11},  are incredibly demanding to implement in a way that closes all loopholes (necessary for security). 

In this work we explore the mid ground, where (local) measurement devices are trusted, but sources and channels are not \cite{PC+12,MP+16,MM15,BM+14}.
 Our statements of confidence are tailored to this end, following the language of quantum authentication \cite{BC+02}, particularly suited to applications for quantum networks. At the end of the protocol one gets a quantum output - the state we want to use  - and a classical output - which tells us weather we accept or reject. A successful test for us is then one that always accepts an ideal source, and outputs the ideal source state (completeness), and if it accepts, the state is not too far from the ideal state (soundness - see below for technical definitions).
 With this in hand, we see how it can be used for certification for various quantum network tasks, in particular for delegated computation, generation of randomness, quantum metrology and quantum secret sharing.

For a given graph $G$ with vertices $V$, and denoting $N(i)$ as the neighbours of $i\in V$, associating a qubit to each vertex, a graph state $|G\rangle$ on $|V|=n$ qubits is defined through the associated stabiliser equations
\begin{eqnarray}
|G\rangle = S_i |G\rangle,
\end{eqnarray}
where $S_i$ are the graph stabiliser operators, with generators $ S_i := X_i \otimes_{j\in N(i)}Z_j$ associated to each of the $N$ veritces, and $X_i$ and $Z_i$ are Pauli operators. We denote the full stabiliser group $S=\{S_i\} = <S_1, ...,S_n>$, which has $2^N$ elements. 
We say that the graph state $|G\rangle$ is shared amongst $n$ players, who depending on the application may be in one physical location or distributed across a network.

The idea of the protocol is very straightforward. The players ask the source for $M$ copies of the graph state. They choose at random one of these to be used, and all the rest are tested by randomly choosing a stabiliser operator and checking it returns the value $+1$. Since the malicious parties (the source, channel... everything except the players) do not know which copy will be tested or used beforehand, the only way they will always pass all the tests is if the players receive the intended graph state each all $M$ times.

\section{Protocol}
Many variants of the protocol are possible, adapted in ways that may depend on the application or implementation at hand.
For clarity we present one particular simple variant of a protocol. After we will comment on other possibilities. 
We start in the standard assumption that the honest parties, the players, share a secret classical key $k=\{r,t\}$,  composed of $r \in [1...M]$, $t=\{t_i\}_{i\ne r}$, $t_i \in [1,..., 2^n]$ denoting $\mathcal{K}$ the set of all keys ($k\in\mathcal{K}$). The protocol follows the steps below.

\begin{enumerate}
	\item The source distributes $M$ $n$-partite systems to the $n$ players. In the honest case, this will be $M$ copies of the graph state $|G\rangle$. \label{PROTOCOL Distribute source}
	\item For copy $i\ne r$, each player performs their part of the measurement of stabiliser $S_{t_i}$. If all the stabilisers output value $+1$, Accept, otherwise Reject.\label{PROTOCOL Stabiliser test}
		\item For copy $r$ the state is the quantum output of the protocol. \label{PROTOCOL Quantum output}
	\end{enumerate}

The variable $M$ plays the role of security parameter (see \eqref{EQN 1/M}).
We briefly comment on some variants. Different parts of the protocol can be changed depending on the application. 
Indeed even the way the secret keys are shared even before the first step may vary, as we will see for the application to secret sharing. 
The way that the outcomes of the test in step \ref{PROTOCOL Stabiliser test} is shared and  acceptance or rejection decided may be important for different cases, for example if some players in the network are dishonest or not trusted.  One may also want to lower the accept threshold in step \ref{PROTOCOL Stabiliser test} to allow for noisy resource states, for example accepting if something less than 100\% of stabiliser tests give the correct output $+1$. We will come back to these variants at different points later, but for now we continue with the simplest version presented above.

\section{Security}

We first formalise our notions of security. For simplicity we encode the classical output as orthogonal quantum states  $|ACC\rangle_R$ for accept and $|REJ\rangle_R$ for reject.
The output state will in general depend on the classical key $k=\{r,t\}$. For each key  $k\in \mathcal{K}$, we denote the output state of the players plus classical reference system as $\rho^k$. 
We say the protocol is $\epsilon$-secure if it satisfies the following two properties
\begin{itemize}
	\item {\bf Completeness}. If the players recieve $M$ copies of the ideal resource state $|G\rangle$, then for all keys $k$ 
	\begin{equation} \label{EQN: completeness}
\rho^k= |G\rangle_P\langle G| \otimes |ACC\rangle_R\langle ACC|.
	\end{equation}
	\item {\bf Soundness}. Denoting the expected output state over all key strings as $\rho_{out} :=\frac{1}{|\mathcal{K}|}\sum_{k\in \mathcal{K}}\rho^k$, and denoting the projection $P_{fail} := (I-|G\rangle_P\langle G|) \otimes |ACC\rangle_R\langle ACC|$, then
	\begin{equation} \label{EQN: Soundness}
	\mathrm{Tr} \left(P_{fail} \rho_{out} \right)\leq \epsilon.
	\end{equation}
\end{itemize}

Completeness is trivially guaranteed since the test uses the stabilisers of the state itself, so it will always accept.
Soundness follows through a similar reasoning to that in \cite{MM15}. Let us denote by $\rho$ the state of all the $M . n$ systems that the players receive in step \ref{PROTOCOL Distribute source} of the protocol.
In order to bound (\ref{EQN: Soundness}) we only need to to consider the output state  conditioned on accept, let us denote it by $\rho_{ACC}$.
To find this we start with the fact that for a given key $k=\{r,t\}$, the projection corresponding to accepting all $M-1$ tests can be written as : 
\begin{equation} \label{M_accept}
	M_{accept}^{r,t} = \bigotimes\limits_{i \neq r} \frac{(S_{t_i} + \mathbb{I}_i)}{2} \otimes \mathbb{I}_r .
\end{equation}
From this we have that 	$\rho_{ACC} $ can be written as 
\begin{equation}
\rho_{ACC} = \sum_{r=1}^M \sum_{t}  \frac{1}{M} \frac{1}{|S|^{M-1}} \rho_{r,t},
\end{equation}
with 
\begin{equation}
\rho_{r,t} = \frac{1}{\mathrm{Tr}(M_{accept}^{r,t}\rho)} \mathrm{Tr}_{ r^c} (M_{accept}^{r,t} \rho) ,
\end{equation}
where $A^c$ denotes the complement of set $A$.

Putting this together, we obtain
\begin{equation}\label{pfailinpartiellespur}
\mathrm{Tr} \left(P_{fail} \rho_{out}\right) = \frac{1}{M} \mathrm{Tr} \left(Q \rho  )\right),
\end{equation}
where
\begin{eqnarray} \label{EQN Q}
Q&=&\sum_{r=1}^M 	\sum_{t}  \frac{1}{S^{M-1}}\bigotimes_{i \neq r} \frac{S_{t_i} + 			\mathbb{I}_i}{2} \otimes \left(\mathbb{I}_r - \left| G \right\rangle \left\langle G 			\right| \right) \nonumber \\
&=& \sum_{r=1}^M \bigotimes_{i \neq r} \frac{\mathbb{I}_i + |G \rangle_i \langle G |}{2} \otimes (\mathbb{I}_r - \left| G \right\rangle_r \left\langle G 	\right|),
\end{eqnarray}
since $1/|S|\sum_i S_i = |G\rangle\langle G|$. Note that  $Q$ is hermitian and positive. It then remains to check that all eigenvalues of $Q$ are smaller than $1$, for which a proof can be found in the appendix. It then follows that 
\begin{equation} \label{EQN 1/M}
\mathrm{Tr}(P_{fail} \rho_{out}) \leq \frac{1}{M} ,
\end{equation}
for all source states $\rho$.

The protocol also has natural extensions for higher prime dimensional graph states, %or continous variable graph states, 
where proofs also follow straightforwardly.
%***** [TRUE for CV?] if so, do in appendix******.

We now present several applications, where the security follows directly as above with a simple application of our protocol, or slight variants of the security statement are made (verified t-designs) or some of the variants of the simplest protocol mentioned above give the utility required (quantum secret sharing).

\section{Applications}

We focus on applications that  can be considered as completely positive trace preserving (CPTP) map $\Gamma$ acting on the quantum output. 
Since fidelity is monotonic under CPTP maps, the usefulness or soundness is preserved. 
This is the case, for example, when further interaction with the source is not required to run the protocol.

Formally, with respect to the CPTP application $\Gamma$ one defines a new fail projector,
\begin{eqnarray}
P_{Fail} (\Gamma):=\left(I- \Gamma \left(|G\rangle \langle G|\right) \right) \otimes |ACC\rangle \langle ACC| .
\end{eqnarray}
Due to the monotonicity of fidelity, (\ref{EQN: Soundness}) implies that 

\begin{equation} \label{EQN: Security Prob Fail CPTP}
Tr \left(P_{fail}^{\Gamma(G)} \Gamma(\rho_B) \right)\leq \frac{1}{M}.
\end{equation}

We now go through some examples of applications.

\subsection{Verified blind quantum computation}
In verified blind quantum computation a technologically limited Alice wishes to delegate some quantum computational task to a server, Bob, in such a way that Bob does not get information about the computation (blind), and moreover, that she can be confident the computation has been carried out correctly (verified).
There are many techniques to achieve this - see \cite{GKK17} for a very recent overview.

In our scenario Alice is limited to single qubit measurements. Clearly this, on its own, is not enough for universal quantum computation. However,  in measurement based quantum computation (MBQC), universal quantum computation is achieved by single qubit measurements on a graph state, with feed forward \cite{RB01}. Importantly the measurements can be made one qubit at a time.
Thus, if Alice asks Bob to provide her with a universal graph states, either cluster states \cite{RB01} or brickwork states~\cite{BFK09l} for example - Alice can perform the computation she wants. Moreover this is blind to Bob - he gets only minimal information, an upper bound to the size of the computation (given by the size of the graph state Alice asks for). To verify the computation Alice can simply apply our protocol to test and use a universal graph state of her choice.

One has the same notions of completeness and soundness as those above, replacing the graph state by the ideal output of the computation. Completeness follows immediately from the universality of the chosen graph state.
For soundness, we simply note that Alice's measurement sequence, which affects the computation, can be understood entirely as a CPTP map on the quantum output of our protocol. 
In this way, the condition (\ref{EQN: Security Prob Fail CPTP}) ensures soundness also. 
More specifically, if we  denote the ideal output of a computation as $\rho_{ideal}^{comp}$, and the average output of a given computation $\rho_{out}^{comp}$, the failing projector becomes $P_{Fail}^{comp}:=\left(I- \rho_{ideal}^{comp} \right) \otimes |ACC\rangle \langle ACC|$, and we have from (\ref{EQN: Security Prob Fail CPTP}) a verification soundness condition (see e.g. \cite{FK12}), 
\begin{equation}
Tr \left(P_{Fail}^{comp} \rho_{out}^{comp} \right)\leq \frac{1}{M}.
\end{equation}
Note that, compared to \cite{FK12}, this scaling with resources is poor. We will talk about this in the conclusions.

We also note that the idea of testing graph states for MBQC computation has been presented before in several measurement based verification schemes, e.g. \cite{HH16}, \cite{HM15}. Indeed, this application of our protocol is almost identical to the verified computation scheme in \cite{HM15}, the main differences being in the specifics of the test (we measure settings chosen from all stabilisers, they a subset) and the figure of merit used (we use the correctness and soundess above, they use the language of hypothesis testing). We present it here simply as an alternative possible scheme, with similar characteristics. As pointed out in \cite{HM15}, this scenario is suited to performing fault tolerant computation, since Alice could equally ask Bob for a resource graph state for fault tolerant computation, for example the topological scheme in \cite{RHG07} using 3D cluster states. This was the idea of the fault tolerant verified computation presented in \cite{FH17}, note however that this works only if the errors on Alice's measurement device are assumed to be independent from anything happening on Bob's side. 

\subsection{Verified t-designs}
Graph states can also be used to sample from a random ensemble of unitaries - this is effectively MBQC without correction, where the measurement outcomes index which unitary is implemented. In particular, in \cite{TM16} it was shown that ensembles with a particularly useful property of being $t$-designs can be efficiently sampled using graph states.
 A $t$-design is an ensemble of unitaries with the property that its statistical moments match those of a Haar ensemble up to order $t$, with applications across quantum information and physics, for example in estimating noise \cite{EW+03}, private channels \cite{HL+04}, modelling thermalisation \cite{MA+15}, photonics\cite{MW+15}, and even black hole physics \cite{HP07}.
Later in \cite{MG+17} this approach was developed to show that efficient  $t$-designs can be generated using a regular lattice similar to the brickwork state. Both results rely heavily on the construction of \cite{BHH12,BHH16} using random circuits.

Our protocol can be used to certify the application of a $t$-design random unitary onto an input, where the source of the graph state is not trusted.
For each set of measurement outcomes $\bar{m}$, we denote the applied CPTP map  on the graph state as $\Gamma^{\bar{m}}$. For simplicity we consider the action of the induced unitary on the  input vertices $I \subset V$ corresponing to inputs in the state $|+\rangle$.
Then \cite{TM16,MG+17} state that measurement result $\bar{m}$, occuring with probability $p_{\bar{m}}$ applies a unitary on the input $|+\rangle^{\otimes|I|}$
\begin{equation} \label{EQN: t-design output}
\Gamma^{\bar{m}} (|G\rangle) = U^{\bar{m}}|+\rangle^{\otimes|I|},
\end{equation}
such that the ensemble $\{p_{\bar{m}}, U^{\bar{m}}\}$ is an approximate $t$-design (see \cite{TM16} for detailed definitions).

For security of verified t-designs one can replace the graph state in the definitions \eqref{EQN: completeness},\eqref{EQN: Soundness} by the output state \eqref{EQN: t-design output}.
The soundness is then guaranteed for each $\bar{m}$ by (\ref{EQN: Security Prob Fail CPTP}). It can easily be seen that one can flip this around to give a statement on the fidelity,
\begin{equation} \label{EQN: Fidelity}
F(U^{\bar{m}}|\psi\rangle,\Gamma^{\bar{m}} (\rho_{ACC}))^2\geq 1-\frac{1}{P_{acc}M},
\end{equation}
where $P_{acc}$ is the probability of passing the tests and $\rho_{ACC}$ is the output of the protocol conditioned on accepting.

\subsection{Quantum Metrology}
In quantum metrology entangled states are used to measure with more precision than is possible with classical probes \cite{GL+11}. The general setting can be understood as an interferometer which imparts a phase $\psi$ on one arm, each time a system passes through it. The idea is to send in many probes $N$ in an entangled state $\rho$, whereafter measurements can reveal the phase with higher precision than possible sending in separable states. 
%Describing the interferometer as a unitary $U_\psi$, we denote the state after it has passed through it as $\rho_\psi = U_\psi^{\otimes n} \rho U_\psi^{\dagger \otimes n}$.

How well this process allows the parameter $\psi$ to be estimated is quantified by the \emph{Quantum Fisher Information}, $\mathcal{F}_Q(\rho)$.
Note, as indicated by the notation, for a simple interferometer the quantum Fisher information is independent of the value of $\psi$ since it is unitarily encoded \cite{TA14,AK+16}. In particular, for $\nu$ independent repetitions of the process, the precision is characterised by the  mean squared error $\Delta^2 \tilde{\psi}$ of a (consistent and unbiased) estimator $\tilde{\psi}$, which is lower bounded by the \emph{Quantum Cram\'{e}r-Rao Bound} \cite{BC94},
\begin{equation}
\Delta^2 \tilde{\psi} \geq \frac{1}{\nu \mathcal{F}_Q(\rho)}.
\end{equation}
For the standard interferometer, the best possible scaling with $N$ is achieved by the $N$-party GHZ state $\frac{1}{\sqrt{2}}(|0\rangle^{\otimes N}+|1\rangle^{\otimes N})$. Denoting its density matrix $\rho_{GHZ}$ we have $\mathcal{F}_Q(\rho_{GHZ})=N^2$. 
The GHZ state is locally equivalent to a graph state for the fully connected graph. Our certification protocol can easily be adapted using the same local unitaries to test $\rho_{GHZ}$ (simply by rotating the test measurements accordingly).

In \cite{AK+16} they show that the  quantum Fisher information of two states differs by an amount bounded by their fidelity
\begin{eqnarray}
\left| \mathcal{F}_Q(\rho)  -  \mathcal{F}_Q(\sigma) \right| \leq 6 \sqrt{1-F(\rho,\sigma)^2}N^2,
\end{eqnarray}
if $\rho$ or $\sigma$ are pure.
That is, if two states are close, as measured by their fidelity, their usefulness for quantum metrology is close. Given the fidelity bound implied by our test (\ref{EQN: Fidelity}), we see that the quantum Fisher information is also bounded. For the rotated protocol testing a GHZ states, given the output state conditioned on accepting $\rho_{ACC}$, we have,
\begin{eqnarray}
\mathcal{F}_Q(\rho_{ACC}) \geq N^2 \left(1-\frac{6}{P_{ACC} M}\right).
\end{eqnarray}

%Considers the case that one does not trust the source of the GHZ states. Our protocol provides a test which then allows us to also be confident of the precision of the estimate.
%In particular, it is clear that, if on performs the same measurements as one would on the $GHZ$ state, the fact that (\ref{EQN: Security Prob Fail CPTP}) ensures the accepted state is not far, implies a bound on the mean squared error by continuity.

 \subsection{Secret sharing over untrusted channels }
 
 In quantum secret sharing a dealer wishes to distribute a secret quantum state amongst $N$ players such that only certain subsets of players can access the secret - the authorised sets. It was shown in \cite{MS08,KF+10} that any secret sharing scheme can be implemented using graph states. However, these rely on the trusted sharing of the graph state. If we are careful, a variant of our protocol can be used to boost these protocols to one where the network of dealer and players do not need to trust the source of the graph state or the channels used to share them. 
 
 There are two important subtleties in the application of our protocol here, stemming from the fact that unauthorised sets of players should be treated as adversaries. Firstly, it makes their inclusion in the stabiliser tests not ideal. Secondly, if they also have access to the random key $k$ this could potentially allow attacks. In \cite{MM15} a protocol was presented which can be understood as a variant of the application of our scheme where i)  the stabiliser tests are restricted to an authorised set, and ii) the classical key $k$ is distributed by a classical secret sharing scheme, with the same access structure. A proof of principle example of this protocol was implemented in \cite{BM+14}, demonstrating its simplicity.
 
% One can think of all secret sharing protocols with $N$ players as encodings of the secret into an $N$ party state. It helps to also think of this as a channel from the dealer to the players. Two features of the code ensure its capacity for secret sharing \cite{MM13}. 
 %First, ensuring that unautorised sets of players do not get information, characterised by low or zero mutual information of the channel to that set of players.
 %Second, that authorised sets of players have full access to the secret - characterised by the fidelity of the output of some decoding operation limited to that set of players. The channel is perfect when the graph state encoding is used. Any interference then acts as noise on this channel and since the channel is independent of the secret, it cannot increase the mutual information of unauthorised sets. Similarly the statement of fidelity follows from the arguments above. 

\section{Conclusions}

In this work we have presented a protocol for certifying graph states and  a few applications in quantum networks.
There are clearly some applications that our protocol would not be suited for - namely ones where further interactions are required with Bob. Such interactions may allow for Bob to correlate his strategy in cheating the `test' part to the future applications potentially threatening functionality (be it security or otherwise).
Nevertheless its simplicity lends itself to many applications as we have seen, not only in the form of protocol presented here, but also its suitability to permit variants, as with secret sharing. 
A simple variant can also deal with noisy states for example, where one would not expect, even an honest noisy source to pass all the time. In such a case one can change the accept requirement to require some smaller portion of correct answers. One can adapt the security statements and proofs to this end without too much difficulty.
%Device independence...
%Graph state self testing (see e.g. applications to quantum computing...)
%Composability
%Fewer measurement choices (simplification of test)

We end with a discussion on scaling of soundness condition  with $S$. 
In the kind of protocol presented here, it is impossible to beat the $1/S$ scaling. This is clear simply because a malicious party can behave honestly for all but one requested state, and send one false/dishonest state. With probability $1/S$ the malicious party's choice of when to be dishonest coincides with the users choice of which one would be used and not tested, so strategy passes the test perfectly yet the state can be arbitrarily far from the ideal one and potentially ruin whatever application. Thus in order to beat the $1/S$ scaling one expects to need some more entanglement. This can be done, for example, by encoding the desired state on some randomly chosen error correcting code - the essential trick used in the original authentication paper by Barnum et al \cite{BC+02}. Such an approach can give an exponential scaling in security with the number of systems sent. The downside now is that the entanglement required scales with the security. 
%Indeed, the protocol in [] can be understood as certifying the EPR state for use in teleportation. 
This then suggests a tradeoff between entanglement and scaling. 

In this context, the advantage of our protocol is that, for many applications, the difficulty in implementing a certified version of an application becomes only the same difficulty as producing the same resource state many times, rather than asking for much more difficult larger, scaling, entanglement.
In optics for example, this advantage makes certified secret sharing possible \cite{BM+14}, doing so an entangled code version would require impractical scaling in entanglement.

\section{Acknowledgements}

We thank Elham Kashefi, Peter Turner and Gérard Duchamp for useful discussions. AK is grateful to the ERASMUS program which supported the visit resulting in this collaboration. DM is grateful for funding from ANR COMB.

\bibliography{biblio}

\newpage
\appendix
\section{Appendix}\label{SEC: App: Bounding Q}

We can write $Q$ (\ref{EQN Q}) as 
\begin{equation}
Q=\sum_{r=1}^M Q_r
\end{equation}

with 
\begin{equation}
Q_r = \bigotimes_{i \neq r} \frac{\mathbb{I}_i + | G \rangle_i \langle G |}{2} \otimes (\mathbb{I}_r - | G \rangle_r \langle G |)
\end{equation}

Let's define $A$ as $\frac{\mathbb{I} + | G \rangle \langle G |}{2}$ and $B$ as $\mathbb{I} - | G \rangle \langle G |$.

An eigenvector for $A$ with eigenvalue $1$ is just given by $|G\rangle$. Denoting $|G'\rangle$ an eigenvector with eigenvalue $1$ of B, an complete eigenbasis for Q is given by all possible combinations of tensor products of those vectors. 
$B$ acts on $| G \rangle$ as $B | G \rangle = | G \rangle - | G \rangle = 0$ while $A | G' \rangle =  \frac{|G'\rangle}{2} + \frac{|G\rangle \langle G | G' \rangle}{2} = \frac{|G' \rangle}{2}$ as $\langle G | G' \rangle = 0$. 

We can then write $Q_r= \bigotimes_{i=1}^{r-1} A_i \bigotimes B_r \bigotimes_{l=r+1}^M A_l$ where $r$ denotes the position of $B$ in the tensor. We then denote the $k$-th family of eigenvectors where $|G'\rangle$ appears $k$ times as ${|Eig \rangle_k = \bigotimes_{j \neq k} |G \rangle_k \bigotimes_{k \neq j} |G 				’\rangle_k | k+j=M}$

Trying to determinate the action from $Q_r$ on $| Eig\rangle_k$, we have to distinguish the following cases :

%\begin{equation}
%Q_r| Eig\rangle_k=(\bigotimes_{i=1}^{r-1} A_i \bigotimes B_r \bigotimes_{l=r+1}^M A_l) \bigotimes_{j=1}^{M-k} |G \rangle \bigotimes_{k=1}^M |G ’\rangle_k
%\end{equation}we have to distinguish the following cases:

Let $r$ be in $ [1, M-k] $. $| G \rangle$ will then be projected to the eigenvalue $0$ so that these cases are trivial. Regarding $r$ in $ [k, M] $ gives us $B |G '\rangle = |G '\rangle$. After then, A acts for $k-1$ times on $| G'\rangle$ giving the eigenvalue $\frac{1}{2^{k-1}}$.

Putting this together and regarding the symmetry of $Q$ with respect to permutations of $r$ within the sum the distribution of $| G \rangle$ and $| G '\rangle$ in the eigenvectors does not matter. It suffices to know, how often $|G ' \rangle$ appears.
The sum contains $M-k$ elements with eigenvalue $0$ and the remaining $k$ elements with eigenvalue $\frac{1}{2^{k-1}}$. The summation gives than as eigenvalue $\frac{k}{2^{k-1}}$ for every $| Eig \rangle_k$ , which is always below one.]

%\section{High dimensions and CV versions}

\end{document}